\documentclass[graybox]{svmult}
\usepackage{graphicx}  
\usepackage{amsmath}   
\usepackage[compress]{cite}
\usepackage{amssymb}   
\usepackage{bm} 
\usepackage{dcolumn}
\usepackage{color}
\usepackage{mathrsfs}
\usepackage{amsfonts}
\usepackage{varioref}
\usepackage{mathptmx}       
\usepackage{helvet}         
\usepackage{courier}        
\usepackage{type1cm}        
\usepackage{makeidx}         
\usepackage{multicol}        
\usepackage[bottom]{footmisc}

\makeindex             

\RequirePackage[colorlinks,citecolor=blue,urlcolor=blue,linkcolor=blue]{hyperref}

\def\EH{Einstein-Hilbert }
\def\LL{Lanczos-Lovelock }

\def\gr{general relativity}

\newcommand{\sqg}{\sqrt{-g}}
\def\GHY{Gibbons-Hawking-York }
\labelformat{section}{Section #1} 
\labelformat{subsection}{Section #1} 
\labelformat{subsubsection}{Section #1}
\labelformat{subsubsubsection}{Section #1}
\labelformat{equation}{Eq.~(#1)} 
\labelformat{figure}{Fig.~#1} 
\labelformat{subfigure}{Fig.~\thefigure#1} 
\labelformat{table}{Table~#1} 
\labelformat{appendix}{Appendix #1}
\begin{document}
\title*{Boundary terms of the Einstein-Hilbert action}
\author{Sumanta Chakraborty}
\institute{Sumanta Chakraborty \at IUCAA, Post Bag 4, Ganeshkhind, Pune University Campus, Pune 411 007, India and
\at Department of Theoretical Physics, Indian Association for the Cultivation of Science, Kolkata 700032, India. \\
\email{sumantac.physics@gmail.com} and \email{sumanta@iucaa.in}}

\maketitle
\abstract*{The Einstein-Hilbert action for general relativity is not well posed in terms of the metric $g_{ab}$ as a dynamical variable. There have been many proposals to obtain an well posed action principle for general relativity, e.g., addition of the Gibbons-Hawking-York boundary term to the Einstein-Hilbert action. These boundary terms are dependent on what one fixes on the boundary and in particular on spacetime dimensions as well. Following recent works of Padmanabhan we will introduce two new variables to describe general relativity and the action principle with these new dynamical variables will turn out to be well posed. Then we will connect these dynamical variables and boundary term obtained thereof to existing literature and shall comment on a few properties of Einstein-Hilbert action which might have been unnoticed earlier in the literature. Before concluding with future prospects and discussions, we will perform a general analysis of the boundary term of Einstein-Hilbert action for null 
surfaces as well.} 
\abstract{The Einstein-Hilbert action for general relativity is not well posed in terms of the metric $g_{ab}$ as a dynamical variable. There have been many proposals to obtain an well posed action principle for general relativity, e.g., addition of the Gibbons-Hawking-York boundary term to the Einstein-Hilbert action. These boundary terms are dependent on what one fixes on the boundary and in particular on spacetime dimensions as well. Following recent works of Padmanabhan we will introduce two new variables to describe general relativity and the action principle with these new dynamical variables will turn out to be well posed. Then we will connect these dynamical variables and boundary term obtained thereof to existing literature and shall comment on a few properties of Einstein-Hilbert action which might have been unnoticed earlier in the literature. Before concluding with future prospects and discussions, we will perform a general analysis of the boundary term of Einstein-Hilbert action for null 
surfaces as well.} 
\section{Introduction}\label{Intro}

Action principle is the starting point of any field theory. Along with the action functional one need to fix the spacetime volume, its boundary and what variable should be fixed on the boundary. When the boundary conditions imposed on an action are compatible with the derived field equation(s), we refer that action principle as well posed. It turns out that the widely used action principle for \gr, the \EH action is \emph{not} well posed. To be more precise, with Ricci scalar as the gravitational Lagrangian, derivation of Einstein's equations requires fixing both metric and its first derivative on the boundary --- inconsistent with Einstein's equations.     

This feature arises, since the action principle for \gr\ is peculiar. It contains second derivatives of the dynamical variables, the metric $g_{ab}$, unlike any other existing Lagrangians. At first glance it seemed quite exotic, since the field equations derived from an action which has second derivatives of the dynamical variable are supposed to have third order derivatives, leading to existence of ghost fields. However it is again the structure of the action principle for \gr\ that comes to rescue. The Ricci scalar can be separated into a bulk term and a surface term. The bulk term has the structure $\Gamma ^{2}$, where $\Gamma ^{a}_{bc}$ are the connection coefficients and along with being quadratic it contains only first derivatives of the metric. In any action principle the surface terms do not contribute to the derivation of field equations, so Einstein's equations also have second derivatives of the metric. However all the second derivatives of the metric hides in the surface term and it is the 
surface term that leads 
to boundary contribution. Hence quite naturally, in the case of \EH action one ends up fixing both the metric and its derivative on the boundary. 

The above arguments pose the problem but also solves it --- it suffices to remove the surface term and consider a new action functional for \gr, namely, $L=R-L_{\rm sur}$, as proposed by Einstein in 1916 \cite{Einstein:1916cd}. Then one obtains Einstein's equations without worrying about the boundary terms. But the problem with the above approach is that, the action is not invariant under diffeomorphism, while we want every action to have the symmetries that the underlying system has. Fortunately, the boundary term that one need to add to the \EH action is by no means unique. Any boundary term that kills all the normal derivatives of the metric on the boundary surface is good enough for our purpose and there could be infinitely many of them as demonstrated by Charap and Nelson in \cite{Charap:1982kn}. The most popular boundary term that keeps the action invariant under diffeomorphism and also makes it well posed is the \GHY term \cite{York:1972sj,Gibbons:1976ue,York1986}. The \GHY term depends on the 
extrinsic curvature $K$ of the boundary surface and is given by $2K\sqrt{|h|}$, where $h$ stands for the determinant of the induced metric on the boundary surface. Note that even though the \GHY boundary term is invariant under diffeomorphism, is not covariant in a strict sense, because of its dependence on the foliation. Further, the \GHY term was guessed and then shown to yield a well posed variational principle without a first principle derivation. This gap was filled by providing a direct derivation of the \GHY boundary term from the action itself in \cite{Padmanabhan:2014BT} while another important issue, the boundary term for null boundaries has been tackled recently in \cite{Parattu:2015gga}. Even then the structure of the boundary term can change depending on what one needs to fix on the boundary, the induced metric or the conjugate momentum and it also changes depending on the spacetime dimensions. For example, it was known that in four-spacetime dimensions if one fix the momentum conjugate to the 
induced metric $h_{ab}$, the \EH action is well posed \cite{York1986,gravitation}. Recently, this result has been explicitly derived in \cite{Krishnan:2016mcj}, but in a broader context and starting from a $D$-dimensional spacetime. 

In this work we will try to provide a broad overview on the possible boundary term structures of the \EH action along with what one needs to fix on the boundary surfaces. This will be performed for both null and non-null cases, besides discussing some other important features of the \EH action.

The paper is organized as follows, in \ref{Sec_02} we will present various boundary terms 
used in various dimensions for an well-posed action of general relativity and their possible connections. Then in \ref{Sec_03} we will explicitly demonstrate some common notions in the context of \gr\ starting from the well known $(1+3)$ decomposition. Finally we comment on the nature of the boundary terms in the context of null surfaces in \ref{Sec_04} before concluding with a brief discussion. We will also present details of the calculations in \ref{PApp_01} and \ref{PApp_02} respectively.

\textit{Notation:} We will work in $D$ spacetime dimensions in \ref{Sec_02}, while the rest of the analysis will be performed in four spacetime dimensions following the mostly positive signature $(-,+,+,+,\ldots)$. The fundamental constants $c$, $G$ and $\hbar$ have been set to unity.
\section{Reconciling boundary terms for the \EH action}\label{Sec_02}

The origin of boundary value problem for \gr\ is due to the fact that \EH action contains second derivatives of the metric --- as a consequence one needs to fix both the metric and its derivatives on the boundary rendering the action ill posed. The above problem arises for using the metric as a fundamental variable and hence to obtain a well posed variational principle we have to add boundary terms to the \EH action. However, it is possible to rewrite the \EH action in the momentum space and the resulting variational principle becomes well posed. The momentum space representation of the \EH action can be obtained by introducing two new variables \cite{Parattu:2013gwb} (see \cite{Chakraborty:2014joa} for a generalization to \LL gravity), 
\begin{align}\label{Eq04}
f^{ab}=\sqrt{-g}g^{ab};\qquad N^{a}_{bc}=Q^{ad}_{be}\Gamma ^{e}_{cd}+Q^{ad}_{ce}\Gamma ^{e}_{bd}=-\Gamma ^{a}_{bc}+\frac{1}{2}\left(\Gamma ^{d}_{bd}\delta ^{a}_{c}+\Gamma ^{d}_{cd}\delta ^{a}_{b}\right)~,
\end{align}
where $f^{ab}$ is a tensor density and $N^{a}_{bc}$ stands for a linear combination of the connections. Note that the above relation holds for any number of spacetime dimensions as $Q^{ab}_{cd}=(1/2)(\delta ^{a}_{c}\delta ^{b}_{d}-\delta ^{a}_{d}\delta ^{b}_{c})$ is independent of spacetime dimensions. However the inverse relation connecting $\Gamma ^{a}_{bc}$ in terms of $N^{a}_{bc}$ depends on the spacetime dimensions and reads in general, 
\begin{align}
\Gamma ^{c}_{ab}=-N^{c}_{ab}+\frac{1}{D-1}\left(N^{d}_{ad}\delta ^{c}_{b}+N^{d}_{bd}\delta ^{c}_{a}\right)~,
\end{align}
which reduces to the expression in \cite{Parattu:2013gwb} for $D=4$. Then the expressions for various curvature components are also modified. For example, the Ricci tensor can be expressed in terms of $N^{c}_{ab}$ such that,
\begin{align}\label{Eq01}
R_{ab}=-\left(\partial _{c}N^{c}_{ab}+N^{c}_{ad}N^{d}_{bc}-\frac{1}{D-1}N^{c}_{ac}N^{d}_{bd}\right)~,
\end{align}
reducing to the one given in \cite{Parattu:2013gwb} for four spacetime dimensions. These variables can be used in the action principle as well, in which case the \EH Lagrangian density becomes $\sqrt{-g}R=f^{ab}R_{ab}$, where $R_{ab}$ can be written in terms of $N^{c}_{ab}$ following \ref{Eq01}. This leads to momentum space representation of the \EH action, which follows from the result that $N^{c}_{ab}=\partial (\sqrt{-g}R)/\partial (\partial _{c}f^{ab})$ and hence the set $(f^{ab},N^{c}_{ab})$ acts as a set of canonically conjugate variables. Further \EH action when varied reads in terms of variations of these canonically conjugate variables as,
\begin{align}
\delta \left(\int _{\mathcal{V}} d^{D}x\sqrt{-g}R\right)&=\int _{\mathcal{V}}d^{D}xR_{ab}\delta f^{ab}-\int _{\mathcal{V}}d^{D}xf^{ab}\nabla _{c}\delta N^{c}_{ab}
\\
&=\int _{\mathcal{V}}d^{D}xR_{ab}\delta f^{ab}-\int _{\partial \mathcal{V}} d^{D-1}x\bar{n}_{c}f^{ab}\delta N^{c}_{ab}~,
\label{Eq02}
\end{align}
where $\mathcal{V}$ stands for the spacetime volume under interest with boundary being denoted by $\partial \mathcal{V}$. 
The last term has been obtained through the use of the following relation $f^{ab}\nabla _{c}\delta N^{c}_{ab}=\partial _{c}\left(\sqrt{-g}g^{ab}\delta N^{c}_{ab}\right)$. Also $\bar{n}_{c}$ in the final expression is the unnormalized normal. If the surface $\partial \mathcal{V}$ is some $\phi =\textrm{constant}$ surface, then $\bar{n}_{c}=\delta ^{\phi}_{c}$. With suitable normalization one obtains, $\bar{n}_{c}=\epsilon (1/N)n_{c}$, where $n_{c}$ is the normalized normal, $\epsilon =\pm 1$ depending on the normal being spacelike or timelike and $N$ is $\sqrt{|g^{\phi \phi}|}$. Thus note that one can obtain the Einstein's equations provided $N^{c}_{ab}$ is fixed at the boundary, leading to an well posed action principle for \gr, since $N^{c}_{ab}$ and $f^{ab}$ are treated as independent variables. 

On the other hand, it is also well known that the variation of the \EH action leads to $\delta (2K\sqrt{h})$, where $K$ is the extrinsic curvature of the boundary surface and $h$ is the determinant of the induced metric on that surface, along with variations of the induced metric with proper coefficients as the boundary term \cite{Padmanabhan:2014BT}. Thus for being consistent one must have the $f^{ab}\delta N^{c}_{ab}$ to yield $\delta (2K\sqrt{h})$ along with variations of the induced metric. It is not at all clear a priori, how this can be achieved. In order to fill this gap we would like to connect the boundary term obtained above in \ref{Eq02} with the standard literature. As a first step towards the connection, we will present a simplified analysis and shall subsequently provide a general derivation.
\subsection{A warm-up example: Analysis in synchronous frame}

Before jumping into the formal derivation let us consider an explicit example as a warm-up. Let us use all the gauge degrees of freedom due to diffeomorphism to eliminate four degrees of freedom from the metric and reduce it to synchronous form, in which the line element reads, 
\begin{align}\label{Eq03}
ds^{2}=-d\tau ^{2}+h_{\alpha \beta}(\tau ,x^{\mu})dx^{\alpha}dx^{\beta}~.
\end{align}
As explicitly demonstrated in \cite{LL2}, any metric can be written in the synchronous coordinate system. The boundary $\partial \mathcal{V}$ of the full spacetime volume can be taken to be $\tau =\textrm{constant}$ hypersurface in this coordinate system, such that the unnormalized normal becomes $\bar{n}_{c}=\delta ^{\tau}_{c}$ and hence the surface term reads,
\begin{align}
\bar{n}_{c}f^{ab}\delta N^{c}_{ab}=f^{ab}\delta N^{0}_{ab}
=-\sqrt{h}\delta N^{0}_{00}+\sqrt{h}h^{\alpha \beta}\delta N^{0}_{\alpha \beta}~,
\end{align}
where in obtaining the last line we have used the synchronous frame metric as in \ref{Eq03}. From the definition of $N^{a}_{bc}$ in terms of connections as in \ref{Eq04} and the metric in \ref{Eq03} it follows that,
\begin{align}
N^{0}_{00}=\Gamma ^{\alpha}_{0\alpha}=-K;\qquad N^{0}_{\alpha \beta}=-\Gamma ^{0}_{\alpha \beta}=K_{\alpha \beta}~.
\end{align}
Thus one can substitute both $N^{0}_{00}$ and $N^{0}_{\alpha \beta}$ in the boundary term which finally leads to, 
\begin{align}
\bar{n}_{c}f^{ab}\delta N^{c}_{ab}&=\sqrt{h}\delta K+\sqrt{h}h^{\alpha \beta}\delta K_{\alpha \beta}
\nonumber
\\
&=\delta \left(2K\sqrt{h}\right)+\sqrt{h}\left(K^{\alpha \beta}-Kh^{\alpha \beta}\right)\delta h_{\alpha \beta}~.
\end{align}
This shows the equivalence of the boundary term with $(f^{ab},N^{c}_{ab})$ as the dynamical variables with the standard boundary term. The above expression explicitly shows that one needs to add $2K\sqrt{h}$ as the boundary term to the \EH action and as a consequence one needs to fix only the spatial part of the metric $h_{\alpha \beta}$ on the boundary $\partial \mathcal{V}$, i.e., on $\tau =\textrm{constant}$ surfaces. 

However the above derivation is a special case and more importantly the boundary term even though is independent of coordinate choices depends heavily on foliation, thus it is not clear from the above result whether the same conclusion should hold for arbitrary foliation as well. This is precisely what we will prove next. 
\subsection{Boundary terms: A general analysis}

As explained above the demonstration in synchronous frame is a specific one among many possible foliations and one needs to provide a general analysis for an arbitrary foliation to grasp the complete structure. To proceed with the general analysis, we will start with the boundary term and shall write $N^{c}_{ab}$ in terms of the connections. Using the fact that variations of the connections are tensors one can ultimately write down the boundary term in terms of the normal and variations in the metric tensor,
\begin{align}\label{Eq05}
\int _{\partial \mathcal{V}} d^{D-1}x\bar{n}_{c}f^{ab}\delta N^{c}_{ab}
=-\int _{\partial \mathcal{V}} d^{D-1}x\bar{n}_{c}\nabla _{d}\left(-\delta g^{cd}+g^{cd}g_{ik}\delta g^{ik}\right)~,
\end{align}
where the following algebraic identity, $-g^{ab}\delta N^{c}_{ab}=\nabla _{d}\left(-\delta g^{cd}+g^{cd}g_{ik}\delta g^{ik}\right)$  have been used in order to arrive at the final result. Given the above \ref{Eq05} we can immediately incorporate the normal inside the covariant derivative and the above expression reads,
\begin{align}
-\int _{\partial \mathcal{V}} d^{D-1}x\bar{n}_{c}f^{ab}\delta N^{c}_{ab}&=\int _{\partial \mathcal{V}}d^{D-1}x\epsilon \sqrt{h}\Bigg\lbrace \nabla _{d}\left(-n_{c}\delta g^{cd}+n^{d}g_{ik}\delta g^{ik}\right)
\nonumber
\\
&-\nabla _{d}n_{c}\left(-\delta g^{cd}+g^{cd}g_{ik}\delta g^{ik}\right)\Bigg\rbrace~,
\end{align}
where $\epsilon =-1$ for spacelike hypersurfaces and is $+1$ for timelike hypersurfaces respectively. The variations of the metric can be divided into two pieces, variations in the induced metric $h_{ij}$ and variations in the normal $n^{i}$. Using the contractions properly and the fact that $\delta (n_{i}n^{i})=0$, we immediately obtain the following expression for the boundary term of the \EH action, 
\begin{align}\label{Eq06}
-\int _{\partial \mathcal{V}} d^{D-1}x\bar{n}_{c}f^{ab}\delta N^{c}_{ab}
&=\int _{\partial \mathcal{V}}d^{D-1}x\epsilon \delta\left(2K\sqrt{h}\right)-\int _{\partial \mathcal{V}}d^{D-1}x\epsilon \sqrt{h}\left(K_{ab}-Kh_{ab}\right)\delta h^{ab}
\nonumber
\\
&+\int d^{D-1}x\epsilon \sqrt{h}D_{i}\left(-n_{c}h^{i}_{b}\delta g^{bc}+2n_{k}h^{i}_{l}\delta g^{kl}\right)~.
\end{align}
The last term is again a surface term and would contribute only on the two surface and hence is neglected. It is useful and instructive to define the momentum conjugate to the induced metric $h_{ab}$ on the hypersurface $\partial \mathcal{V}$ as,
\begin{align}
\Pi _{ab}=\sqrt{h}\left(K_{ab}-Kh_{ab}\right)~.
\end{align}
Note that $n_{a}\Pi ^{ab}=0$. Thus finally using the expression for $\Pi _{ab}$ and neglecting the surface term, we obtain the simplified version of the boundary term from \ref{Eq06} in the most general case as, 
\begin{align}\label{Eq07}
-\int _{\partial \mathcal{V}} d^{D-1}x\bar{n}_{c}f^{ab}\delta N^{c}_{ab}&=\int _{\partial \mathcal{V}}d^{D-1}x\epsilon \delta\left(2K\sqrt{h}\right)-\int _{\partial \mathcal{V}}d^{D-1}x\epsilon \Pi _{ab}\delta h^{ab}~.
\end{align}
The result in the synchronous frame can be derived immediately from the above relation by substituting $\epsilon =-1$, since $\tau =\textrm{constant}$ surfaces are spacelike. However note that the two-dimensional surface terms identically vanishes in the synchronous frame. The above result suggests that if we add the boundary term $-2\epsilon K\sqrt{h}$ to the \EH action the normal derivatives of the metric will be removed from the boundary and one needs to fix only the induced metric $h^{ab}$. It is important to emphasis at this stage that fixing $h^{ab}$ is different from fixing $h_{ab}$. Since by construction we have $n_{a}\propto \nabla _{a}\phi$, and $n_{a}h^{ab}=0$, this suggests $h^{ab}=h^{\alpha \beta}$, where $\alpha ,\beta$ are spacetime indices excluding $\phi$, while $h_{ab}$ has all the metric components. Due to the momentum and Hamiltonian constraints of \gr\ one cannot fix all the metric components on the hypersurfaces and hence the correct variational principle would be the one which fixes 
only $h^{ab}$, i.e., $h^{\alpha \beta}$ on the boundary $\partial \mathcal{V}$. 

Let us now illustrate the fact that $2\epsilon K\sqrt{h}$ is not the only boundary term that can lead to a well-posed action principle for \gr, there are infinitely many. However for our illustration we will pick two of them. Since we are working in a $D$ dimensional spacetime we have the following identity, $\Pi _{ab}h^{ab}= -(D-2)K\sqrt{h}$. We can use the above identity to convert the original result in \ref{Eq07} to two different results,
\begin{align}\label{Eq08}
-\int _{\partial \mathcal{V}} d^{D-1}x\bar{n}_{c}f^{ab}\delta N^{c}_{ab}&=\int _{\partial \mathcal{V}}d^{D-1}x\epsilon \delta\left(2K\sqrt{h}\right)-\int _{\partial \mathcal{V}}d^{D-1}x\epsilon \delta\left(\Pi _{ab}h^{ab}\right)
\nonumber
\\
&+\int _{\partial \mathcal{V}}d^{D-1}x\epsilon h^{ab} \delta \Pi _{ab}
\nonumber
\\
&=\int _{\partial \mathcal{V}}d^{D-1}x\epsilon \delta\left(DK\sqrt{h}\right)
+\int _{\partial \mathcal{V}}d^{D-1}x\epsilon h^{ab} \delta \Pi _{ab}~.
\end{align}
The above result depicts that one can also add $-D\epsilon K\sqrt{h}$ as the boundary term to the \EH action and hence obtain an well-posed variational principle if $\Pi _{ab}$ is fixed at the boundary. Note that as we have argued earlier, the only non-zero components of $h^{ab}$ are $h^{\alpha \beta}$ and hence one need to fix only $\Pi _{\alpha \beta}$ at the boundary $\partial \mathcal{V}$. This result can also be casted in a different form, for that we need to use the identity, $\Pi _{ab}\delta h^{ab}=-\Pi ^{ab}\delta h_{ab}$. Use of which enables one to write \ref{Eq07} in the following form 
\begin{align}\label{Eq09}
-\int _{\partial \mathcal{V}} d^{D-1}x\bar{n}_{c}f^{ab}\delta N^{c}_{ab}&=\int _{\partial \mathcal{V}}d^{D-1}x\epsilon \delta\left(2K\sqrt{h}\right)+\int _{\partial \mathcal{V}}d^{D-1}x\epsilon \delta\left(\Pi ^{ab}h_{ab}\right)
\nonumber
\\
&-\int _{\partial \mathcal{V}}d^{D-1}x\epsilon h_{ab} \delta \Pi ^{ab}
\nonumber
\\
&=\int _{\partial \mathcal{V}}d^{D-1}x\epsilon \delta\left[\left(4-D\right)K\sqrt{h}\right]
-\int _{\partial \mathcal{V}}d^{D-1}x\epsilon h_{ab} \delta \Pi ^{ab}~.
\end{align}
This is another form of the boundary contribution, recently derived in \cite{Krishnan:2016mcj} but from a different perspective. In our case, the result essentially follows from the original boundary term, expressed in terms of the canonically conjugate variables $(f^{ab},N^{c}_{ab})$. In this case the boundary term one has to add to the \EH action corresponds to, $(4-D)\epsilon K\sqrt{h}$, with the peculiarity that at $D=4$ this term identically vanishes. While in this case one need to fix $\Pi ^{ab}$ at the boundary $\partial \mathcal{V}$. Hence the original boundary term from which all possible versions of the boundary terms including the well-known $2\epsilon K\sqrt{h}$ can be derived is the $f^{ab}\delta N^{c}_{ab}$ combination. Further we have shown two explicit examples in which one can add different boundary term at the expense of fixing either $\Pi ^{ab}$ or $\Pi _{ab}$ at the boundary (see \ref{table_01}). Even though it is tempting to assume $h_{ab}\delta \Pi ^{ab}=-h^{ab}\delta \Pi _{ab}$, this relation is actually 
not correct. This can be seen from the following algebraic 
manipulation straightforwardly,  
\begin{align}
h^{ab}\delta \Pi _{ab}&=h^{ab}\delta \left(h_{ac}h_{bd}\Pi ^{cd}\right)=h_{cd}\delta \Pi ^{cd}+2\Pi ^{ac}\delta h_{ac}
\nonumber
\\
&=h_{cd}\delta \Pi ^{cd}-2\Pi _{ac}\delta h^{ac}=-h_{cd}\delta \Pi ^{cd}+\delta \left[\left(4-2D\right)K\sqrt{h}\right]~,
\end{align}
reconciling the two results presented in \ref{Eq08} and \ref{Eq09} respectively. Through this exercise we have achieved two important goals, which are,
\begin{svgraybox}
\begin{itemize}

\item By introducing the canonically conjugate variables $(f^{ab},N^{c}_{ab})$, one obtains the Einstein's equations from variations of $f^{ab}$, while variations of $N^{c}_{ab}$ leads to the boundary term. Hence the \EH action becomes action in the momentum space such that one need to fix the momentum $N^{c}_{ab}$ at the boundary. However there were no clear consensus how this boundary term is related to the existing ones, e.g., the \GHY boundary term. In this section we have explicitly demonstrated the connection, by deriving the \GHY counter term starting from the boundary term consisting of $f^{ab}\delta N^{c}_{ab}$. 

\item Secondly, in most of the literatures people always take the \GHY boundary term to be the only boundary term possible. In the last part of this section we have explicitly demonstrated two more boundary terms. Our result clearly shows that the structure of the boundary term depends crucially on what one fixes at the boundary. If one fixes the induced metric $h^{ab}$, then \GHY term is the only option. But if one fixes the conjugate momentum, then depending on whether one fixes $\Pi ^{ab}$ or $\Pi _{ab}$, one arrives to different boundary terms. In particular when $\Pi ^{ab}$ is fixed one need not add any boundary term in four dimensions, which is a peculiar feature of \gr. 

\end{itemize}
\end{svgraybox}
Thus we have reconciled the possible boundary terms that one can add to the \EH action. Their non-uniqueness and derivation from a first principle starting from \EH action in momentum space has also been presented. We will now turn to the $(1+3)$ decomposition of the \EH action and related comments. 
\begin{table}
\caption{A comparison of various boundary terms of \EH action}
\label{table_01}       
%
%
\begin{tabular}{p{1.5cm}p{2.5cm}p{2.5cm}p{2cm}p{2.5cm}}
\hline\noalign{\smallskip}
Bulk & Surface & Boundary  & What to fix    & Well-Posed \\
Term & Term    & Term $^{a}$     & on Boundary    & Action \\
\noalign{\smallskip}\svhline\noalign{\smallskip}
$R_{ab}\delta f^{ab}$ & $-\bar{n}_{c}f^{ab}\delta N^{c}_{ab}$  & None                               & $N^{c}_{ab}$  & $\sqrt{-g}R$ \\
\\
$G_{ab}\delta g^{ab}$ & $\epsilon \delta (2K\sqrt{h})$         & $\epsilon \delta (2K\sqrt{h})$     & $h^{ab}$      &  
$\sqrt{-g}R-\epsilon \delta (2K\sqrt{h})$ \\
                      & $-\epsilon \Pi _{ab}\delta h^{ab}$     &                                    &               &              \\
\\
$G_{ab}\delta g^{ab}$ & $\epsilon \delta (DK\sqrt{h})$         & $\epsilon \delta (DK\sqrt{h})$     &  $\Pi _{ab}$  &  
$\sqrt{-g}R-\epsilon \delta (DK\sqrt{h})$  \\
                      & $\epsilon h^{ab}\delta \Pi _{ab}$      &                                    &               &           
                      \\
\\
$G_{ab}\delta g^{ab}$ & $\epsilon \delta [(4-D)K\sqrt{h}]$     & $\epsilon \delta [(4-D)K\sqrt{h}]$ & $\Pi ^{ab}$   & 
$\sqrt{-g}R$\\
                      &  $-\epsilon h_{ab}\delta \Pi ^{ab}$    &                                    &               &
$-\epsilon \delta [(4-D)K\sqrt{h}]$                \\
\noalign{\smallskip}\hline\noalign{\smallskip}
\end{tabular}
$^{a}$ Note that in the last case for $D=4$ no boundary term is needed and \EH action is well posed, with $\Pi ^{ab}$ fixed on the boundary (see also \cite{Krishnan:2016mcj}).
\end{table}
\section{(1+3) decomposition, time derivatives and canonical momenta}\label{Sec_03}

In \gr\ space and time are treated on an equal footing. However for many application, e.g., canonical quantization schemes, one need the notion of time and hence the splitting of four dimensional spacetime into one time and three spatial coordinates becomes immediate. This has been performed successfully by Arnowitt, Deser and Misner (henceforth referred to as ADM) in a seminal work \cite{Arnowitt:1962hi}, in which the ten independent metric components are split into three pieces --- $h_{\alpha \beta}$, $N^{\alpha}$ and $N$, such that, the line element becomes
\begin{equation}\label{ADM_01}
ds^{2}=-N^{2}dt^{2}+h_{\alpha \beta}\left(dx^{\alpha}+N^{\alpha}dt\right)\left(dx^{\beta}+N^{\beta}dt\right)~.
\end{equation} 
Thus note that the spatial metric $g_{\alpha \beta}$ is just $h_{\alpha \beta}$, the off-diagonal entries are $N_{\alpha}\equiv h_{\alpha \beta}N^{\beta}$, while the temporal component of the metric becomes, $g_{00}=-N^{2}+h_{\alpha \beta}N^{\alpha}N^{\beta}$. For the inverse metric the temporal component is simple but not the spatial components such that,
\begin{equation}\label{ADM_02}
g^{tt}=-\frac{1}{N^{2}},\qquad g^{t\alpha}=\frac{N^{\alpha}}{N^{2}},\qquad 
g^{\alpha \beta}=\left(h^{\alpha \beta}-\frac{N^{\alpha}N^{\beta}}{N^{2}}\right)~.
\end{equation}
The next point one can address from the ADM splitting corresponds to the $(1+3)$ decomposition of the \EH action. This would require projection of the Riemann tensor components on the spacelike hypersurface, leading to $^{(3)}R$, the Ricci scalar of the spacelike hypersurface and invariants like $K_{ab}K^{ab}$, $K^{2}$ constructed out of the extrinsic curvature components \cite{gravitation,Arnowitt:1962hi}
\begin{align}\label{Eq11}
\sqrt{-g}R&=\sqrt{-g}\left[^{(3)}R+K_{ab}K^{ab}-K^{2}-2\nabla _{i}\left(Kn^{i}+a^{i}\right) \right]
\nonumber
\\
&=\sqrt{-g}L_{\rm ADM}-2\sqrt{-g}\nabla _{i}\left(Kn^{i}+a^{i}\right)~,
\end{align}
where $n_{i}$ is the normal to the spacelike hypersurface and $a^{i}$ is the corresponding acceleration. Thus the \EH Lagrangian can be written in terms of the ADM Lagrangian and an additional boundary term which coincides with the \GHY counter term since $n_{i}a^{i}=0$. It is well known that the ADM Lagrangian does not contain time derivatives of $N$ and $N^{\alpha}$ and hence their conjugate momentums vanish. Thus these variables are non-dynamical. However we have just witnessed that boundary terms are not unique, one can in principle add any boundary term that cancels the normal derivative. Then a natural question arises --- are the time derivatives of $N$ and $N^{\alpha}$ zero for for any possible boundary term? If not can they be dynamical? These questions get firm ground as the following example is considered. 
\begin{svgraybox}
\begin{center}
\textbf{Dynamical or Non-dynamical?}
\end{center}
\vskip 0.5 true cm
Let us consider a cosmological spacetime. Being homogeneous and isotropic it is described by a single function, the scale factor $a(t)$. The line element for cosmological spacetime by imposition of these symmetry conditions become,
\begin{align}
ds^{2}=-dt^{2}+a^{2}(t)\left[dr^{2}+r^{2}d\Omega ^{2}\right]~,
\end{align}
where the spatial section has been assumed to be flat for simplicity. The above metric is manifestly in ADM form, with $N=1$, $N^{\alpha}=0$ and $h_{\alpha \beta}=a^{2}(t)\delta _{\alpha \beta}$ respectively. Thus it is evident that $N$ and $N^{\alpha}$ are not dynamical, all the dynamics comes from the scale factor $a(t)$ as expected. One can now introduce a new coordinate $r$, such that $R=a(t)r$ and write the metric in the $(t,R,\theta,\phi)$ coordinate system such that,
\begin{align}
ds^{2}=-\left(1-H^{2}R^{2}\right)dt^{2}-2HRdtdR+dR^{2}+R^{2}d\Omega ^{2}~.
\end{align}
Surprisingly, now the metric is again in ADM form but with a completely different structure. This time the spatial metric is flat, i.e., $h_{\alpha \beta}=\delta _{\alpha \beta}$ and hence cannot have any dynamics. On the other hand, one obtains $N=1$ and $N^{\alpha}=HR\delta ^{\alpha}_{R}$ and would conclude that cosmological spacetime is non-dynamical! This explicitly shows that the standard argument for ADM variables $N$ and $N^{\alpha}$ to be non-dynamical based on their time derivatives is misleading.
\end{svgraybox}
To resolve the dilemma we will explicitly illustrate, depending on the boundary term, \EH action do contains time derivatives of $N$ and $N^{\alpha}$ but they are \emph{not} dynamical. For this purpose we make use of the following decomposition of the \EH action, 
\begin{align}\label{Eq10}
\sqrt{-g}R=\sqrt{-g}g^{ab}\left(\Gamma ^{i}_{ja}\Gamma ^{j}_{ib}-\Gamma ^{i}_{ab}\Gamma ^{j}_{ij}\right)+\partial _{c}\left\lbrace \sqrt{-g}\left(g^{ik}\Gamma ^{c}_{ik}-g^{ck}\Gamma ^{m}_{km}\right)\right\rbrace~. 
\end{align}
Here the first term is quadratic in the connection and is known as the $\Gamma ^{2}$ Lagrangian, while the second term is the boundary term and contains normal derivatives of the metric as elaborated in \cite{gravitation}. Thus an alternative to \GHY boundary term is the total divergence term introduced above and hence a possible well-posed Lagrangian corresponds to the $\Gamma ^{2}$ Lagrangian. We will show that this Lagrangian depends on time derivatives of $N$ and $N^{\alpha}$. To achieve this we shall expand out the $\Gamma ^{2}$ Lagrangian in terms of the ADM variables and separate out the time derivatives of $N$ and $N^{\alpha}$. Any term $X$ which contains time derivatives of $N$ and $N^{\alpha}$ will be denoted by $[X]_{t.d}$. By Expressing all the connections in terms of the ADM variables we find that only $\Gamma ^{t}_{tt}$ and $\Gamma ^{\alpha}_{tt}$ depends on time derivatives of $N$ and $N^{\alpha}$. Hence the time derivative part for the full $\Gamma ^{2}$ Lagrangian reads (see \ref{PApp_01} 
for detailed derivation),
\begin{align}\label{gamma2}
\left[\sqrt{-g}L_{\rm quad}\right]_{t.d}= \frac{\sqrt{h}}{N^{2}}\partial _{t}N\partial _{\alpha}N^{\alpha}
-\sqrt{h}\frac{\partial _{t}N^{\alpha}\partial _{\alpha}N}{N^{2}}
+\frac{\partial _{t}N^{\alpha}}{N}\partial _{\alpha}\sqrt{h}~.
\end{align}
Hence we have explicitly demonstrated, that the $\Gamma ^{2}$ Lagrangian contains time derivatives of $N$ and $N^{\alpha}$. Then one question naturally arises, how is that the ADM Lagrangian does not contain these time derivative terms, as evident from the expression for $L_{\rm ADM}$? The answer to this question is hiding in the boundary terms, since they are not identical. Thus in order to understand this, we will have to compare the two boundary terms, the surface term in \ref{Eq10} and the \GHY boundary term, that separate $\Gamma^2$ Lagrangian and ADM Lagrangian, respectively, from the Einstein-Hilbert Lagrangian $\sqg R$.

Let us now evaluate the Einstein and the \GHY boundary terms using the ADM variables. We shall not evaluate the integrands of the surface integrals, but the corresponding divergence terms present in the bulk Lagrangians given by \ref{Eq11} and \ref{Eq10} respectively. One can again use the Christoffel symbols to calculate $K n^i+a^i$ required for evaluating the \GHY term in divergence form. Performing the same, terms in the \GHY boundary contribution containing time derivatives of $N$ and $N^{\alpha}$ has the expression
\begin{align}
\Bigg[-2\partial _{i}\left\lbrace \sqrt{-g}\left(Kn^{i}+a^{i}\right)\right\rbrace\Bigg]_{t.d}
&=\sqrt{h}\frac{\partial _{t}N\partial _{\alpha}N^{\alpha}}{N^{2}}
-2\sqrt{h}\frac{\partial _{t}\partial _{\alpha}N^{\alpha}}{N}
-2\frac{\partial _{t}\sqrt{h}\partial _{t}N}{N^{2}}
\nonumber
\\
&+2\frac{\partial _{t}N}{N^{2}}N^{\alpha}\partial _{\alpha}\sqrt{h}
-2\frac{\partial _{t}N^{\alpha}\partial _{\alpha}\sqrt{h}}{N}~. \label{Kt}
\end{align}
Having derived the relevant expressions related to \GHY boundary term, let us next concentrate on the boundary term in the \EH action given in \ref{Eq10}, which has the expression $\partial _{i}(\sqrt{-g}V^{i})$, where $V^{i}=g^{ab}\Gamma ^{i}_{ab}-g^{im}\Gamma ^{k}_{mk}$. Computation of each individual components of the boundary term which contains time derivatives of $N$ and $N^{\alpha}$ are thus given by
\begin{align}\label{Vtd}
\left[\partial _{i}\left(\sqrt{-g}V^{i}\right)\right]_{t.d}=&-\frac{2}{N^{2}}\partial _{t}N\partial _{t}\sqrt{h}
+\frac{\sqrt{h}}{N^{2}}\partial _{t}N\partial _{\alpha}N^{\alpha}
-\frac{\sqrt{h}}{N}\partial _{t}\partial _{\alpha}N^{\alpha}
+\frac{2}{N^{2}}\partial _{t}N N^{\alpha}\partial _{\alpha}\sqrt{h}
\nonumber
\\
&-\frac{2}{N}\partial _{t}N^{\alpha}\partial _{\alpha}\sqrt{h}
+\frac{\sqrt{h}}{N^{2}}\partial _{\alpha}N\partial _{t}N^{\alpha}
-\frac{\sqrt{h}}{N}\partial _{\alpha}\partial _{t}N^{\alpha}
-\frac{\partial _{t}N^{\alpha}\partial _{\alpha}\sqrt{h}}{N}~.
\end{align}
Hence, from \ref{Kt} and \ref{Vtd}, we finally arrive at the total contribution from the boundary terms
\begin{align}\label{surf_term_diff}
\Bigg[\partial _{c}\left(\sqrt{-g}V^{c}\right)&+2\partial _{i}\left\lbrace \sqrt{-g}\left(Kn^{i}+a^{i}\right)\right\rbrace \Bigg]_{t.d}
\nonumber
\\
&=-\frac{\sqrt{h}}{N^{2}}\partial _{t}N \partial _{\alpha}N^{\alpha}
+\sqrt{h}\frac{\partial _{t}N^{\alpha}\partial _{\alpha}N}{N^{2}}
-\frac{\partial _{t}N^{\alpha}\partial _{\alpha}\sqrt{h}}{N}~.
\end{align}
Thus, we observe that the surface terms in Einstein-Hilbert action in Einstein's original decomposition and ADM decomposition are different. The difference contains time derivatives of $N^{\alpha}$ and $N$. These time derivatives should exactly match the time derivatives in $\Gamma^2$ Lagrangian as we know that the ADM Lagrangian does not have time derivatives of $N$ and $N^{\alpha}$. Evaluating time derivatives in ADM Lagrangian using \ref{gamma2} and \ref{surf_term_diff}, we obtain
\begin{align}
\left[\sqrt{-g}L_{\rm ADM}\right]_{t.d}&=\left[\sqrt{-g}R + 2\partial _{i}\left\lbrace \sqrt{-g} \left(Kn^{i}+a^{i}\right)\right\rbrace\right]_{t.d} 
\nonumber
\\
&=\left[\sqrt{-g}L_{\rm quad}
+\partial _{c}\left(\sqrt{-g}V^{c}\right)+2\partial _{i}\left\lbrace \sqrt{-g}\left(Kn^{i}+a^{i}\right)\right\rbrace \right]_{t.d}
\nonumber
\\
&=0~,
\end{align}
which confirms the ADM Lagrangian does not contain any time derivatives of $N$ and $N^{\alpha}$ and demonstrates that the time derivatives of $N$ and $N^{\alpha}$ in the $\Gamma^2$ action arise because of the difference in surface terms.

Since the $\Gamma ^{2}$ Lagrangian contains time derivatives of $N$ and $N^{\alpha}$, it is pertinent to ask what are the conjugate momenta corresponding to $N$ and $N^{\alpha}$. From \ref{gamma2}, the conjugate momenta for $N$ and $N^{\alpha}$ turn out to be
\begin{align}\label{can_mom_N_Na}
p_{(N)}&=\frac{\partial \left(\sqrt{-g}\Gamma ^{2}\right)}{\partial \left(\partial _{t}N\right)}=\frac{\sqrt{h}}{N^{2}}\partial _{\alpha}N^{\alpha}
\\
p_{\alpha ~(N^{\alpha})}&=\frac{\partial \left(\sqrt{-g}\Gamma ^{2}\right)}{\partial \left(\partial _{t}N^{\alpha}\right)}=-\sqrt{h}\frac{\partial _{\alpha}N}{N^{2}}
+\frac{1}{N}\partial _{\alpha}\sqrt{h}~.
\end{align}
Note that the conjugate momenta to $N$ and $N^{\alpha}$ do not depend on time derivatives of $N$ and $N^{\alpha}$ respectively. Hence, these relations cannot be inverted to obtain $\partial _{t}N$ and $\partial _{t}N^{\alpha}$ in terms of $p_{(N)}$ and $p_{\alpha ~(N^{\alpha})}$. Returning back to our example of cosmological spacetime, this means that $H$ is indeed non-dynamical and that is clear since in terms of Hubble parameter, the Einstein's equations contain only single time derivative of $H$. Thus we conclude:
\begin{svgraybox}
Even though the ADM Lagrangian does not contain time derivatives of $N$ and $N^{\alpha}$, the quadratic Lagrangian $L_{\rm quad}$ differing from the ADM Lagrangian by total derivative do contains time derivatives of $N$ and $N^{\alpha}$. However, the corresponding canonical momentums are non-invertible, i.e., one cannot obtain time derivatives of $N$ and $N^{\alpha}$ in terms of their canonical momentum. Hence follows their non-dynamical nature.
\end{svgraybox}
This explicitly demonstrates standard statements, showing truth in non-dynamical behavior of $N$ and $N^{\alpha}$ but also demonstrating existence of time derivatives of non-dynamical variables.
\section{Null Surfaces: completing the circle}\label{Sec_04}

The boundary terms and ADM decomposition discussed earlier depends crucially on the timelike (or spacelike) nature of the boundary surface. However, the most ubiquitous surfaces in \gr\ are the null surfaces, e.g., in a black hole spacetime the standard boundary would consist of the surface at infinity and the event horizon, which is a null surface. The limit of non-null surfaces to null surfaces is not at all straightforward, since many quantities including the extrinsic curvature, induced metric can either blow up or vanish on the null surface if proper care is not taken. Thus it is important to consider the boundary term from a first principle in connection to null hypersurfaces. The first step towards this direction was taken in \cite{Padmanabhan:2014BT} by constructing a general formalism and its explicit implementation was carried out in \cite{Parattu:2015gga}. There it was argued that for a null vector $\ell _{a}$ (i.e., $\ell ^{a}\ell _{a}=0$) the boundary term one should add corresponds to 
$2\sqrt{q}(\Theta +\kappa)$, where $q$ stands for the determinant of the induced metric on the null surface, $\Theta$ stands for the expansion of the null geodesics and $\kappa$ is the non-affinity parameter. Since null surfaces are intrinsically two-dimensional, use of a single vector field $\ell _{a}$ is not sufficient. One need to introduce another auxiliary vector field $k_{a}$, satisfying $k_{a}k^{a}=0$ and $\ell _{a}k^{a}=-1$. In the above derivations it has been assumed that the null surface is preserved under variations, i.e., the following three conditions hold: $\delta (\ell _{a}\ell ^{a})=0$, $\delta (\ell _{a}k^{a})=0$ and finally $\delta (k_{a}k^{a})=0$. In this work we will relax all these assumptions and shall investigate the effect of these constraints on the boundary term and degrees of freedom on the boundary. We will start with the general expression for boundary term of \EH action having the form \cite{Parattu:2016trq}
\begin{align}\label{Eq1}
\sqrt{-g}Q[\ell _{c}]&=\sqrt{-g}\nabla _{c}\left(\delta u^{c}\right)-2\delta \left(\sqrt{-g}\nabla _{a}\ell ^{a}\right)+\sqrt{-g}\left[\nabla _{a}\ell _{b}-g_{ab}\left(\nabla _{c}\ell ^{c}\right)\right]\delta g^{ab}
\nonumber
\\
&=Q_{1}+Q_{2}+Q_{3}~,
\end{align}
where, $\delta u^{a}=\delta \ell ^{a}+g^{ab}\delta \ell _{b}$. Note that the above expression yields boundary term of the \EH action for $\ell _{c}=\nabla _{c}\phi$. Since we are interested in null vectors only, there can be one more parametrization of null vectors, namely, $\ell _{c}=A\nabla _{c}\phi$. In this case the boundary term would be, $(1/A)\sqrt{-g}Q[\ell _{c}]$.

We have separated the $\sqrt{-g}Q[\ell _{c}]$ term in \ref{Eq1} into three natural combinations, one is a divergence term, $Q_{1}$, second one corresponds to total variation $Q_{2}$ and finally the degrees of freedom term $Q_{3}$ respectively. We will explore each of these terms and subsequently shall evaluate the boundary term on the null surface following the convention, if some relation holds \emph{only} on the null surface it will be denoted by $A:=0$. As explained above we will assume the following conditions on the null surface only, $\ell _{a}\ell ^{a}:=0$, $\ell _{a}k^{a}:=0$ and $k_{a}k^{a}:=0$ respectively, but we would not assume anything about off the null surface relations, i.e., variations can be arbitrary. Then one can introduce the \emph{partial} projector $P^{a}_{~b}$ through the vectors $\ell ^{a}$ and $k^{a}$ as, $P^{a}_{~b}=\delta ^{a}_{b}+k^{a}\ell _{b}$ (which satisfies the relation $\ell _{a}P^{a}_{b}=0$) and can write the first divergence term $Q_{1}$ in \ref{Eq1} as:
\begin{align}
Q_{1}:&=\partial _{\alpha} \left(\sqrt{-g}P^{\alpha}_{~d}\delta u^{d}\right)-\delta \left(\sqrt{-g}k^{c}\partial _{c}\ell ^{2} \right)+\left(k^{c}\partial _{c}\ell ^{2}\right)\delta \sqrt{-g}
\nonumber
\\
&+\sqrt{-g}\delta k^{c}\partial _{c}\ell ^{2}
-\partial _{c}\left( \sqrt{-g}k^{c}\right)\delta \ell ^{2}~,
\end{align}
while the second term can also be expressed using the partial projector $P^{a}_{~b}$ and then the complete $\sqrt{-g}Q[\ell _{c}]$ term on using the variation of $\sqrt{-g}$, takes the following form
\begin{align}\label{Eq1a}
\sqrt{-g}Q\left[\ell _{c}\right]&:=\partial _{\alpha} \left(\sqrt{-g}P^{\alpha}_{~d}\delta u^{d}\right)
-2\delta \left(\sqrt{-g}P^{a}_{~b}\nabla _{a}\ell ^{b}\right)
+\sqrt{-g}\delta k^{c}\partial _{c}\ell ^{2}
\nonumber
\\
&-\partial _{c}\left(\sqrt{-g}k^{c}\right)\delta \ell ^{2}+\sqrt{-g}\left(\nabla _{a}\ell _{b}-g_{ab}\left\lbrace P^{c}_{~d}\nabla _{c}\ell ^{d} \right\rbrace \right)\delta g^{ab}~.
\end{align}
Note that the first term is a pure surface term --- it has no component along the normal $\ell _{a}$. Then we can decompose the metric in terms of the induced metric $q_{ab}$ and the null vectors $\ell ^{a}$ and $k^{a}$ as: $g_{ab}=q_{ab}-\ell _{a}k_{b}-\ell _{b}k_{a}$. Thus variations of the metric now gets transformed to variations of the induced metric and the null vectors. One important point to keep in mind is the fact that $\delta \ell ^{a}=g^{ab}\delta \ell _{b}+\ell _{b}\delta g^{ab}$ but \emph{not} $g^{ab}\delta \ell _{b}$. Using the properties of the null vectors outside variation and decomposition of $\nabla _{a}\ell _{b}$ in terms of the extrinsic curvature ultimately lands us into the following expression for the $\sqrt{-g}Q[\ell _{c}]$ term (see \ref{PApp_02} for a derivation of this result)
\begin{align}\label{Final_Eq}
\sqrt{-g}Q[\ell _{c}]&:=\partial _{\alpha}\left(\sqrt{-g}P^{\alpha}_{~a}\delta u^{a}\right)
-2\delta \left(\sqrt{-g}P^{a}_{~b}\nabla _{a}\ell ^{b}\right)
\nonumber
\\
&+\sqrt{-g}\left[\Theta _{ab}-\left(P^{c}_{~d}\nabla _{c}\ell ^{d}\right)q_{ab}\right]\delta q^{ab}
\nonumber
\\
&-\sqrt{-g}\left\lbrace k^{m}\nabla _{m}\ell _{a}+k^{n}\nabla _{a}\ell _{n}
+\left(k^{m}k^{n}\nabla _{m}\ell _{n}\right)\ell _{a}-2\left(P^{c}_{~d}\nabla _{c}\ell ^{d}\right)k_{a} \right\rbrace \delta \ell ^{a}
\nonumber
\\
&+\sqrt{-g}\left\lbrace k^{m}\nabla _{m}\ell ^{a}+k^{n}\nabla ^{a}\ell _{n}
+\left(k^{m}k^{n}\nabla _{m}\ell _{n}\right)\ell ^{a} -2\left(P^{m}_{~n}\nabla _{m}\ell ^{n}\right)k^{a} \right\rbrace \delta \ell _{a}
\nonumber
\\
&+\sqrt{-g}\left\lbrace \partial _{c}\ell ^{2}\right\rbrace \delta k^{c}-\partial _{c}\left(\sqrt{-g}k^{c}\right)\delta \ell ^{2}~,
\end{align}
which is the boundary term for $\ell _{c}=\nabla _{c}\phi$. There exist another simpler expression for $Q[\ell _{c}]$ alone, without the $\sqrt{-g}$ factor. Sometimes this expression also becomes useful and hence we state the final result here (details can be found in \ref{PApp_02})
\begin{align}\label{Final_Eq_Sim}
Q[\ell _{c}]&=\nabla _{c}\delta u^{c}-2\delta \left(\nabla _{c}\ell ^{c}\right)+\nabla _{a}\ell _{b}\delta g^{ab}
\nonumber
\\
&:=\frac{1}{\sqrt{-g}}\partial _{\alpha}\left(\sqrt{-g}P^{\alpha}_{~a}\delta u^{a}\right)
-2\delta \left(P^{a}_{~b}\nabla _{a}\ell ^{b}\right)+\Theta _{ab}\delta q^{ab}
\nonumber
\\
&-\left\lbrace k^{m}\nabla _{m}\ell _{a}+k^{n}\nabla _{a}\ell _{n}
+\left(k^{m}k^{n}\nabla _{m}\ell _{n}\right)\ell _{a}\right\rbrace \delta \ell ^{a}
+\left\lbrace \partial _{c}\ell ^{2}\right\rbrace \delta k^{c}
\nonumber
\\
&+\left\lbrace k^{m}\nabla _{m}\ell ^{a}+k^{n}\nabla ^{a}\ell _{n}
+\left(k^{m}k^{n}\nabla _{m}\ell _{n}\right)\ell ^{a}\right\rbrace \delta \ell _{a}
-\frac{1}{\sqrt{-g}}\partial _{c}\left(\sqrt{-g}k^{c}\right)\delta \ell ^{2}~.
\end{align}
On the other hand, if $\ell _{c}=A\nabla _{c}\phi$, the boundary term on the null surface becomes,
\begin{align}\label{Final_Eq_A}
\frac{\sqrt{-g}}{A}Q[\ell _{c}]&:=\partial _{\alpha}\left(\frac{\sqrt{-g}}{A}P^{\alpha}_{~a}\delta u^{a}\right)
-2\delta \left(\frac{\sqrt{-g}}{A}P^{a}_{~b}\nabla _{a}\ell ^{b}\right)
\nonumber
\\
&+\frac{\sqrt{-g}}{A}\left[\Theta _{ab}-\left(P^{c}_{~d}\nabla _{c}\ell ^{d}\right)q_{ab}\right]\delta q^{ab}
-\frac{\sqrt{-g}}{A}\Big\lbrace k^{m}\nabla _{m}\ell _{a}+k^{n}\nabla _{a}\ell _{n}
\nonumber
\\
&+\left(k^{m}k^{n}\nabla _{m}\ell _{n}\right)\ell _{a}-2\left(P^{c}_{~d}\nabla _{c}\ell ^{d}\right)k_{a} 
-\nabla _{a}\ln A \Big\rbrace \delta \ell ^{a}+\frac{\sqrt{-g}}{A}\Big\lbrace k^{m}\nabla _{m}\ell ^{a}
\nonumber
\\
&+k^{n}\nabla ^{a}\ell _{n}
+\left(k^{m}k^{n}\nabla _{m}\ell _{n}\right)\ell ^{a} -2\left(P^{m}_{~n}\nabla _{m}\ell ^{n}\right)k^{a}+\nabla ^{a}\ln A \Big\rbrace \delta \ell _{a}
\nonumber
\\
&+\frac{\sqrt{-g}}{A}\left\lbrace \partial _{c}\ell ^{2}\right\rbrace \delta k^{c}-\partial _{c}\left(\frac{\sqrt{-g}}{A}k^{c}\right)\delta \ell ^{2}-2\frac{\sqrt{-g}}{A}\left(\Theta +\kappa \right)\delta \ln A~.
\end{align}
Before commenting on the structure of the boundary term let us quickly check two possible limits that have been derived earlier in \cite{Parattu:2015gga}. The first one corresponds to the situation in which $\ell _{a}=\nabla _{a}\phi$. Assuming the hypersurfaces to be unchanged due to variation, one obtains $\delta \ell _{a}=0$. Further imposing the following conditions: $\delta \ell ^{2}=0=\delta \left(\ell ^{a}k_{a}\right)$, and using the result $P^{a}_{~b}\nabla _{a}\ell ^{b}=\Theta +\kappa$, the boundary term can be reduced to:
\begin{align}\label{Derived_01}
\sqrt{-g}Q\left[\nabla _{c}\phi\right]&:=\partial _{\alpha}\left(\sqrt{-g}P^{\alpha}_{~a}\delta u^{a}\right)
-2\delta \left[\sqrt{-g}\left(\Theta +\kappa \right)\right]
\nonumber
\\
&+\sqrt{-g}\left[\Theta _{ab}-\left(\Theta +\kappa\right)q_{ab}\right]\delta q^{ab}
\nonumber
\\
&-2\sqrt{-g}\left\lbrace k^{m}\nabla _{m}\ell _{a}
-\left(\Theta +\kappa\right)k_{a} \right\rbrace \delta \ell ^{a}~.
\end{align}
This is exactly what have been obtained by various other routes in \cite{Parattu:2015gga}. The second result corresponds to similar limit of \ref{Final_Eq_A}, i.e., the variations keep the hypersurfaces unchanged and hence, $\delta \ell _{a}=(\delta \ln A)\ell _{a}$. Further we assume, $\delta (\ell ^{2})=0$ and $\delta (\ell ^{a}k_{a})=0$, which helps to reduce \ref{Final_Eq_A} to the following form,
\begin{align}\label{Derived_02}
\frac{\sqrt{-g}}{A}Q[\ell _{c}]&:=\partial _{\alpha}\left(\frac{\sqrt{-g}}{A}P^{\alpha}_{~a}\delta u^{a}\right)
-2\delta \left[\frac{\sqrt{-g}}{A}\left(\Theta +\kappa\right)\right]
\nonumber
\\
&+\frac{\sqrt{-g}}{A}\left[\Theta _{ab}-\left(\Theta +\kappa\right)q_{ab}\right]\delta q^{ab}
\nonumber
\\
&+\frac{\sqrt{-g}}{A}\Bigg\lbrace 2\left(\Theta +\kappa\right)k_{a}-k^{m}\left(\nabla _{m}\ell _{a}+\nabla _{a}\ell _{m}\right)
+\nabla _{a}\ln A \Bigg\rbrace \delta \ell ^{a}~.
\end{align}
This one also coincides exactly with the result derived in \cite{Parattu:2015gga}.

Having checked the consistency with earlier derived results we now concentrate on the physical implications of \ref{Final_Eq}. The first term as emphasized earlier corresponds to another boundary term \footnote{This kind of terms are also present in the the calculation for spacelike (or timelike) surfaces, see for example the last term of \ref{Eq06}.} and contributes only on the two surface without much significance. The second term is the boundary term that one should add (negative of that term, to be precise) to the \EH action when evaluated within a volume enclosed by null boundaries. The rest of the terms tells us what one should fix on the null surface. Among them fixing induced metric is expected, with its conjugate momentum being $\pi _{ab}=\sqrt{-g}\left[\Theta _{ab}-\left(P^{m}_{~n}\nabla _{m}\ell ^{n}\right)q_{ab}\right]$. In this case as well one can write $\pi _{ab}\delta q^{ab}$ term as a total variation leading to a different boundary term and conjugate momentum to fix on the boundary. However, 
unlike the cases of timelike or spacelike surfaces the situation is not so simple for null surfaces, since even after fixing the induced metric one needs to fix the components of the null vectors as well. But one can improve on that. Since the normalization of the null vector is arbitrary one can always choose $\ell _{a}$ to be a pure gradient such that $\delta \ell _{a}=0$. Further, since the choice of $k^{a}$ is arbitrary one might chose it such that it satisfies $(1/\sqrt{-g})\partial _{c}(\sqrt{-g}k^{c})=0$, as well as $\delta (k^{a}\ell _{a})=0$. Then, one obtains, $\delta k^{c}\partial _{c}\ell ^{2}\propto \ell _{c}\delta k^{c}=0$. As these seemingly natural conditions are being satisfied, the boundary term simplifies a lot, ultimately leading to,
\begin{svgraybox}
\begin{align}
\sqrt{-g}Q[\ell _{c}]&:=\partial _{\alpha}\left(\sqrt{-g}P^{\alpha}_{~a}\delta u^{a}\right)
-2\delta \left(\sqrt{-g}P^{a}_{~b}\nabla _{a}\ell ^{b}\right)
\nonumber
\\
&+\sqrt{-g}\left[\Theta _{ab}-\left(P^{c}_{~d}\nabla _{c}\ell ^{d}\right)q_{ab}\right]\delta q^{ab}
\nonumber
\\
&-\sqrt{-g}\left\lbrace k^{m}\nabla _{m}\ell _{a}+k^{n}\nabla _{a}\ell _{n}
+\left(k^{m}k^{n}\nabla _{m}\ell _{n}\right)\ell _{a}-2\left(P^{c}_{~d}\nabla _{c}\ell ^{d}\right)k_{a} \right\rbrace \delta \ell ^{a}~.
\end{align}
\end{svgraybox}
Hence along with $q_{ab}$ one need to fix the components of the null vector $\ell ^{a}$. One more point should be noted, since $\delta \ell _{a}=0$, one obtains $\delta (\ell _{a}\ell ^{a})=\ell _{a}\delta \ell ^{a}$ and hence any contribution from $\delta \ell ^{2}$ can be dumped into the contribution from $\delta \ell ^{a}$. This suggests that on the null surface one need to fix the induced metric $q^{ab}$ as well as $\ell ^{a}$, having interesting consequences for degrees of freedom on the null surfaces \`{a} la degrees of freedom on spacelike or timelike surfaces. One interesting corollary could be, as the diffeomorphisms are gauged away, one can eliminate the four degrees of freedom in $\delta \ell ^{c}$, keeping the true (physical) degrees of freedom in the two metric $q_{ab}$ of the null surface. This can have interesting implications for black hole entropy, which we will pursue elsewhere. 
\section{Concluding Remarks}

The peculiarity of the \EH action can be traced back to its boundary terms. In the standard treatments it is often overlooked that \EH action is not well posed, one has to add boundary terms to get an well posed action for gravity. There have been parallel results on this issue, one is the well-known \GHY boundary term, while the other is recent and more promising from a thermodynamic hindsight which invokes two new variables $f^{ab}$ and $N^{c}_{ab}$ to describe gravity, with $f^{ab}\delta N^{c}_{ab}$ as the boundary term. In this work we have explicitly derived the equivalence between these two formalisms in any spacetime dimensions. Further we have also demonstrated the argument that ``boundary terms are not unique'' by constructing two more boundary terms starting from the \GHY term. To our surprise these boundary terms depend strongly on the spacetime dimensions and even can vanish in $D=4$. Then we have elaborated the meaning of another statement made often in the literature, ``the ADM variables $N$ 
and $N^{\alpha}$ are not dynamical''. The standard argument goes by saying that the ADM Lagrangian does not depend on time derivatives of $N$ and $N^{\alpha}$. We have shown that one can add boundary terms to the ADM Lagrangian leading to a new Lagrangian which contain time derivatives of $N$ and $N^{\alpha}$, (so it might appear they can be made dynamical by adding boundary terms) but still they are non-dynamical as conjugate momentums to them cannot be inverted. This finishes our discussion on spacelike or timelike surfaces and we turn to the case of null surfaces. In earlier works regarding boundary term on null surfaces, various assumption about variations of the null vectors were imposed, here we have derived the structure of the boundary term for most general variation. Imposing some minimal restrictions, we have shown that besides the induced metric, the null vector $\ell ^{a}$ contains additional degrees of freedom. If they can be removed by diffeomorphism (as \cite{Parattu:2015gga} suggests) then 
the induced metric might contain all the physical degrees of freedom associated with null surfaces, having implications for emergent paradigms of gravity \cite{Padmanabhan:2003gd,Padmanabhan:2009vy,Padmanabhan:2014jta,Chakraborty:2015hna,
Chakraborty:2015wma,Chakraborty:2015aja,Chakraborty:2014rga}. 
\section*{Acknowledgements}

Research of S.C. is funded by a SPM fellowship from CSIR, Government of India. He thanks T. Padmanabhan for suggesting this problem and for comments on an earlier version of the manuscript. He also thanks J. Bagla and S. Engineer for invitation to write this article along with K. Parattu, S. Singh and K. Lochan for interesting discussions and comments. A part of this work was completed during a visit to ICTP, Italy and the author thanks ICTP for warm hospitality.
\section*{Dedication}

One of the great quality of Prof. Padmanabhan is his ability to ask the correct question. This work stems from such questions asked by him during our discussions: ``What is the connection between various action principles for \gr? Why $N$ and $N^{\alpha}$ are non-dynamical? What one should fix on a null surface?''. I have tried to answer them in this work and I respectfully dedicate it to Prof. Padmanabhan on the occasion of his $60$th birthday. 
\appendix
\labelformat{section}{Appendix #1} 
\labelformat{subsection}{Appendix #1}
\section{Appendix: Time dependence of $N$ and $N^{\alpha}$}\label{PApp_01}

This appendix is devoted in filling the gaps in the derivation of time dependence of ADM variables $N$ and $N^{\alpha}$ presented in the main text. For completeness, we will also present Christoffel connections for the ADM decomposition as they will be heavily required for obtaining time dependence of $N$ and $N^{\alpha}$ respectively. For the metric presented in \ref{ADM_01} and the inverse metric in \ref{ADM_02} the Christoffel symbols have the following expressions:
\begin{align}
\Gamma ^{t}_{tt}&=\frac{\partial _{t}N}{N}-\frac{N_{\alpha}N_{\beta}}{2N^{2}}\partial _{t}h^{\alpha \beta}
-\frac{N_{\beta}N_{\mu}}{2N^{2}}N^{\alpha}\partial _{\alpha}h^{\beta \mu}
+\frac{1}{N}N^{\alpha}\partial _{\alpha}N-\frac{1}{N^{2}}N^{\alpha}N^{\beta}\partial _{\alpha}N_{\beta}~,
\end{align}
\begin{align}
\Gamma ^{t}_{\alpha t}&=\frac{\partial _{\alpha}N}{N}-\frac{N^{\beta}\partial _{\alpha}N_{\beta}}{N^{2}}
-\frac{N_{\beta}N_{\mu}}{2N^{2}}\partial _{\alpha}h^{\beta \mu}
+\frac{N^{\beta}}{2N^{2}}\partial _{t}h_{\alpha \beta}
+\frac{N^{\beta}}{2N^{2}}\left(\partial _{\alpha}N_{\beta}-\partial _{\beta}N_{\alpha}\right)~,
\end{align}
\begin{align}
\Gamma ^{t}_{\alpha \beta}&=\frac{1}{2N^{2}}\left[\partial _{t}h_{\alpha \beta}
-\left(\partial _{\alpha}N_{\beta}+\partial _{\beta}N_{\alpha}\right)\right]
+\frac{N^{\mu}}{2N^{2}}\left(-\partial _{\mu}h_{\alpha \beta}+\partial _{\alpha}h_{\beta \mu}
+\partial _{\beta}h_{\alpha \mu} \right)~,
\end{align}
\begin{align}
\Gamma ^{\alpha}_{tt}&=-\frac{N^{\alpha}\partial _{t}N}{N}
+h^{\alpha \beta}\left(N\partial _{\beta}N-N^{\mu}\partial _{\beta}N_{\mu}
-\frac{1}{2}N_{\mu}N_{\nu}\partial _{\beta}h^{\mu \nu}+\partial _{t}N_{\beta}\right)
\nonumber
\\
&-\frac{N^{\alpha}N^{\beta}}{N^{2}}\left(N\partial _{\beta}N-N^{\mu}\partial _{\beta}N_{\mu}
-\frac{1}{2}N_{\mu}N_{\nu}\partial _{\beta}h^{\mu \nu} \right)+\frac{N^{\alpha}N_{\mu}N_{\nu}\partial _{t}h^{\mu \nu}}{2N^{2}}~,
\end{align}
\begin{align}
\Gamma ^{\alpha}_{t\beta}&=-\frac{N^{\alpha}\partial _{\beta}N}{N}
+\frac{N^{\alpha}N^{\mu}\partial _{\beta}N_{\mu}}{N^{2}}
+\frac{N^{\alpha}}{2N^{2}}N_{\mu}N_{\nu}\partial _{\beta}h^{\mu \nu}
\nonumber
\\
&+\frac{1}{2}\left(h^{\alpha \mu}-\frac{N^{\alpha}N^{\mu}}{N^{2}}\right)\left(-\partial _{\mu}N_{\beta}+\partial _{\beta}N_{\mu}
+\partial _{t}h_{\mu \beta} \right)~,
\end{align}
\begin{align}
\Gamma ^{\alpha}_{\mu \beta}&=\Gamma ^{(3)\alpha}_{\mu \beta}
+\frac{N^{\alpha}}{2N^{2}}\left(-\partial _{t}h_{\mu \beta}+\partial _{\mu}N_{\beta}
+\partial _{\beta}N_{\mu}\right)
\nonumber
\\
&-\frac{N^{\alpha}N^{\nu}}{2N^{2}}\left(-\partial _{\nu}h_{\mu \beta}+\partial _{\beta}h_{\mu \nu}
+\partial _{\mu}h_{\beta \nu}\right)~.
\end{align}
Having derived the relevant connections we will now proceed to derive the time dependent terms in the $\Gamma ^{2}$ Lagrangian. 
\subsection{$\Gamma ^{2}$ Lagrangian}

In this section, we shall expand out the $\Gamma^2$ Lagrangian in terms of the ADM variables and separate out the time derivatives of $N$ and $N^{\alpha}$. As already mentioned in the main text, any term $X$ which contains time derivatives of $N$ and $N^{\alpha}$ will be denoted by $[X]_{t.d}$. We start with the the expression for the $\Gamma ^{2}$ Lagrangian given by
\begin{equation}\label{Gamma2Lag}
L_{\rm quad}=g^{ab}\left(\Gamma ^{i}_{ja}\Gamma ^{j}_{ib}-\Gamma ^{i}_{ab}\Gamma ^{j}_{ij}\right)~.
\end{equation}
Let us start with the second term in \ref{Gamma2Lag} having the expression
\begin{align}
-g^{ab}\Gamma ^{i}_{ab}\Gamma ^{j}_{ij}&=-g^{ab}\Gamma ^{i}_{ab}\partial _{i}\left(\ln \sqrt{-g}\right)
\nonumber
\\
&=-g^{ab}\Gamma ^{i}_{ab}\partial _{i}\left[\ln \left(N\sqrt{h}\right)\right]
\nonumber
\\
&=\frac{1}{N^{2}}\Gamma ^{i}_{00}\partial _{i}\left[\ln \left(N\sqrt{h}\right)\right]
-2\frac{N^{\alpha}}{N^{2}}\Gamma ^{i}_{0\alpha}\partial _{i}\left[\ln \left(N\sqrt{h}\right)\right]
\nonumber
\\
&-\left(h^{\alpha \beta}-\frac{N^{\alpha}N^{\beta}}{N^{2}}\right)\Gamma ^{i}_{\alpha \beta}
\partial _{i}\left[\ln \left(N\sqrt{h}\right)\right]~.
\end{align}
Now we will retain only those terms which can potentially contain time derivatives of $N$ and $N^{\alpha}$. From the above expression, such terms turn out to have the following expression:
\begin{align} 
\left[-g^{ab}\Gamma ^{i}_{ab}\Gamma ^{j}_{ij}\right]_{t.d}
&=\frac{1}{N^{2}}\left(\frac{\partial _{t}N}{N}+\frac{\partial _{t}\sqrt{h}}{\sqrt{h}}\right)
\Gamma ^{t}_{tt}+\frac{1}{N^{2}}
\left(\frac{\partial _{\alpha}N}{N}+\frac{\partial _{\alpha}\sqrt{h}}{\sqrt{h}}\right)\Gamma ^{\alpha}_{tt}
\nonumber
\\
&-2\frac{N^{\alpha}}{N^{3}}\Gamma ^{t}_{t\alpha}\partial _{t}N -\left(h^{\alpha \beta}-\frac{N^{\alpha}N^{\beta}}{N^{2}}\right)\Gamma ^{t}_{\alpha \beta}
\frac{\partial _{t}N}{N}~.
\end{align} 
The first term in \ref{Gamma2Lag} can be decomposed as
\begin{align}
g^{ab}\Gamma ^{i}_{ja}\Gamma ^{j}_{ib}&=-\frac{1}{N^{2}}\Gamma ^{i}_{jt}\Gamma ^{j}_{it}
+2\frac{N^{\alpha}}{N^{2}}\Gamma ^{i}_{jt}\Gamma ^{j}_{i\alpha}
+\left(h^{\alpha \beta}-\frac{N^{\alpha}N^{\beta}}{N^{2}}\right)\Gamma ^{i}_{j\alpha}\Gamma ^{j}_{i\beta}~.
\end{align}
Again retaining only those terms that contain time derivatives of $N$ and $N^{\alpha}$, we finally arrive at
\begin{align}
\left[g^{ab}\Gamma ^{i}_{ja}\Gamma ^{j}_{ib}\right]_{t.d}
=-\frac{1}{N^{2}}\left(\Gamma ^{t}_{tt}\Gamma ^{t}_{tt}+2\Gamma ^{t}_{\alpha t}\Gamma ^{\alpha}_{tt}\right)
+\frac{N^{\alpha}}{N^{2}}\left(\Gamma ^{t}_{tt}\Gamma ^{t}_{t\alpha}
+\Gamma ^{\beta}_{tt}\Gamma ^{t}_{\beta \alpha}\right)~.
\end{align}
Thus, the terms that involve time derivatives of $N$ and $N^{\alpha}$ in the $\Gamma ^{2}$ Lagrangian turn out to be
\begin{align}
\left[L_{\rm quad}\right]_{t.d}&=\frac{1}{N^{2}}\left[\Gamma ^{t}_{tt}\left(\frac{\partial _{t}N}{N}
+\frac{\partial _{t}\sqrt{h}}{\sqrt{h}}-\Gamma ^{t}_{tt}\right)
+\Gamma ^{\alpha}_{tt}\left(\frac{\partial _{\alpha}N}{N}+\frac{\partial _{\alpha}\sqrt{h}}{\sqrt{h}}
-2\Gamma ^{t}_{\alpha t}\right)\right]
\nonumber
\\
&+2\frac{N^{\alpha}}{N^{2}}\Gamma ^{\beta}_{tt}\Gamma ^{t}_{\alpha \beta}
-\left(h^{\alpha \beta}-\frac{N^{\alpha}N^{\beta}}{N^{2}}\right)\Gamma ^{t}_{\alpha \beta}
\frac{\partial _{t}N}{N}
\nonumber
\\
&=\frac{1}{N^{2}}\Bigg[\Gamma ^{t}_{tt}\left(\frac{\partial _{t}N}{N}
+\frac{\partial _{t}\sqrt{h}}{\sqrt{h}}-\Gamma ^{t}_{tt}\right)
\nonumber
\\
&+\Gamma ^{\alpha}_{tt}\left(\frac{\partial _{\alpha}N}{N}+\frac{\partial _{\alpha}\sqrt{h}}{\sqrt{h}}
-2\Gamma ^{t}_{\alpha t}+2N^{\beta}\Gamma ^{t}_{\alpha \beta}\right)\Bigg]
-\left(h^{\alpha \beta}-\frac{N^{\alpha}N^{\beta}}{N^{2}}\right)\Gamma ^{t}_{\alpha \beta}
\frac{\partial _{t}N}{N}~.
\end{align}
The first term above can be manipulated as follows:
\begin{align}
\frac{1}{N^{2}}\Gamma ^{t}_{tt}\left(\frac{\partial _{t}N}{N}
+\frac{\partial _{t}\sqrt{h}}{\sqrt{h}}-\Gamma ^{t}_{tt}\right)
&=\frac{1}{N^{2}}\Gamma ^{t}_{tt}\Big(\frac{\partial _{t}N}{N}
+\frac{\partial _{t}\sqrt{h}}{\sqrt{h}}-\frac{\partial _{t}N}{N}
+\frac{N_{\alpha}N_{\beta}}{2N^{2}}\partial _{t}h^{\alpha \beta}
\nonumber
\\
&+\frac{N_{\beta}N_{\mu}}{2N^{2}}N^{\alpha}\partial _{\alpha}h^{\beta \mu}
-\frac{1}{N}N^{\alpha}\partial _{\alpha}N+\frac{1}{N^{2}}N^{\alpha}N^{\beta}\partial _{\alpha}N_{\beta}\Big)
\nonumber
\\
&=\frac{\partial _{t}N}{N^{3}}\Big(\frac{\partial _{t}\sqrt{h}}{\sqrt{h}}
+\frac{N_{\alpha}N_{\beta}}{2N^{2}}\partial _{t}h^{\alpha \beta}
\nonumber
\\
&+\frac{N_{\beta}N_{\mu}}{2N^{2}}N^{\alpha}\partial _{\alpha}h^{\beta \mu}
-\frac{1}{N}N^{\alpha}\partial _{\alpha}N+\frac{1}{N^{2}}N^{\alpha}N^{\beta}\partial _{\alpha}N_{\beta}\Big)~.
\end{align}
We also have the following identity:
\begin{align}
-2\Gamma ^{t}_{\alpha t}&+2N^{\beta}\Gamma ^{t}_{\alpha \beta}
=-2\frac{\partial _{\alpha}N}{N}+2\frac{N^{\beta}\partial _{\alpha}N_{\beta}}{N^{2}}
+\frac{N_{\beta}N_{\mu}}{N^{2}}\partial _{\alpha}h^{\beta \mu}
-\frac{N^{\beta}}{N^{2}}\partial _{t}h_{\alpha \beta}
\nonumber
\\
&-\frac{N^{\beta}}{N^{2}}\left(\partial _{\alpha}N_{\beta}-\partial _{\beta}N_{\alpha}\right)
+\frac{N^{\beta}}{N^{2}}\left[\partial _{t}h_{\alpha \beta}
-\left(\partial _{\alpha}N_{\beta}+\partial _{\beta}N_{\alpha}\right)\right]
+\frac{N^{\beta}N^{\mu}}{N^{2}}\left(\partial _{\alpha}h_{\beta \mu}\right)
\nonumber
\\
&=-2\frac{\partial _{\alpha}N}{N}~.
\end{align}
Substituting all these expressions, we finally arrive at
\begin{align}
\Big[&L_{\rm quad}\Big]_{t.d}=\frac{\partial _{t}N}{N^{3}}\Bigg(\frac{\partial _{t}\sqrt{h}}{\sqrt{h}}
+\frac{N_{\alpha}N_{\beta}}{2N^{2}}\partial _{t}h^{\alpha \beta}
+\frac{N_{\beta}N_{\mu}}{2N^{2}}N^{\alpha}\partial _{\alpha}h^{\beta \mu}-\frac{1}{N}N^{\alpha}\partial _{\alpha}N
\nonumber
\\
&+\frac{1}{N^{2}}N^{\alpha}N^{\beta}\partial _{\alpha}N_{\beta}\Bigg)+\frac{1}{N^{2}}\Gamma ^{\alpha}_{tt}\left(-\frac{\partial _{\alpha}N}{N}
+\frac{\partial _{\alpha}\sqrt{h}}{\sqrt{h}}\right)
-\left(h^{\alpha \beta}-\frac{N^{\alpha}N^{\beta}}{N^{2}}\right)\Gamma ^{t}_{\alpha \beta}
\frac{\partial _{t}N}{N}
\nonumber
\\
&=\frac{\partial _{t}N^{\alpha}}{N^{2}}\left(-\frac{\partial _{\alpha}N}{N}
+\frac{\partial _{\alpha}\sqrt{h}}{\sqrt{h}}\right)
+\frac{\partial _{t}N}{N^{3}}\Big(\frac{\partial _{t}\sqrt{h}}{\sqrt{h}}
+\frac{N_{\alpha}N_{\beta}}{2N^{2}}\partial _{t}h^{\alpha \beta}
+\frac{N_{\beta}N_{\mu}}{2N^{2}}N^{\alpha}\partial _{\alpha}h^{\beta \mu}
\nonumber
\\
&-\frac{1}{N}N^{\alpha}\partial _{\alpha}N
+\frac{1}{N^{2}}N^{\alpha}N^{\beta}\partial _{\alpha}N_{\beta}
+\frac{N^{\alpha}\partial _{\alpha}N}{N}
-\frac{N^{\alpha}\partial _{\alpha}\sqrt{h}}{\sqrt{h}}
-\left(h^{\alpha \beta}-\frac{N^{\alpha}N^{\beta}}{N^{2}}\right)N^{2}\Gamma ^{t}_{\alpha \beta}\Big)~.
\end{align}
Separating out only the time derivatives of $N$ and $N^{\alpha}$, we obtain \ref{gamma2} as presented in the main text. 
As we have demonstrated, the $\Gamma^2$ Lagrangian contains time derivatives of $N$ and $N^{\alpha}$. Let us now evaluate the boundary terms to complete the circle.
\subsection{The Boundary Terms}

Let us now use these Christoffel symbols to compute $(K n^i+a^i)$ required for evaluating the \GHY counter-term in divergence form. The time component of the acceleration $a^i$ vanishes, thanks to the relation $n_{i}a^{i}=0$, while the spatial components are non-zero and are given by the expression
\begin{align}
a^{\alpha}&=\frac{1}{N}\partial _{t}\left(-\frac{N^{\alpha}}{N}\right)
+\frac{N^{\beta}}{N}\partial _{\beta}\left(\frac{N^{\alpha}}{N}\right)
+\frac{1}{N^{2}}\Gamma ^{\alpha}_{tt}-2\frac{N^{\beta}}{N^{2}}\Gamma ^{\alpha}_{t\beta}
+\frac{N^{\beta}N^{\mu}}{N^{2}}\Gamma ^{\alpha}_{\mu \beta}
\nonumber
\\
&=-\frac{N_{\beta}\partial _{t}h^{\alpha \beta}}{N^{2}}
+\frac{N^{\beta}N_{\mu}\partial _{\beta}h^{\alpha \mu}}{N^{2}}
+\frac{N^{\beta}N^{\mu}}{N^{4}}\Gamma ^{(3)\alpha}_{\mu \beta}
+\frac{\partial _{\beta}N}{N}h^{\alpha \beta}
\nonumber
\\
&-\frac{N_{\mu}N_{\nu}}{2N^{4}}h^{\alpha \beta}\partial _{\beta}h^{\mu \nu}
-\frac{N^{\mu}}{N^{2}}h^{\alpha \beta}\partial _{t}h_{\beta \mu}~.
\end{align}
The extrinsic curvature $K$ can be calculated to be
\begin{align}
K&=-\nabla _{i}n^{i}=-\partial _{i}n^{i}-n^{i}\partial _{i}\ln \sqrt{-g}
\nonumber
\\
&=-\partial _{t}\left(\frac{1}{N}\right)+\partial _{\alpha}\left(\frac{N^{\alpha}}{N}\right)
-\frac{1}{N}\partial _{t}\left[\ln \left(N\sqrt{h}\right)\right]
+\frac{N^{\alpha}}{N}\partial _{\alpha}\left[\ln \left(N\sqrt{h}\right)\right]
\nonumber
\\
&=\frac{\partial _{\alpha}N^{\alpha}}{N}-\frac{1}{N}\partial _{t}\left(\ln \sqrt{h}\right)
+\frac{N^{\alpha}}{N}\partial _{\alpha}\left(\ln \sqrt{h}\right)~.
\end{align}
Therefore, we obtain the components of the vector $A^{i}=Kn^{i}+a^{i}$ as follows:
\begin{align}
A^{t}&=\frac{\partial _{\alpha}N^{\alpha}}{N^{2}}-\frac{1}{N^{2}}\partial _{t}\left(\ln \sqrt{h}\right)
+\frac{N^{\alpha}}{N^{2}}\partial _{\alpha}\left(\ln \sqrt{h}\right)
\nonumber
\\
A^{\alpha}&=a^{\alpha}
-\frac{N^{\alpha}}{N}
\left(\frac{\partial _{\alpha}N^{\alpha}}{N}-\frac{1}{N}\partial _{t}\left(\ln \sqrt{h}\right)
+\frac{N^{\alpha}}{N}\partial _{\alpha}\left(\ln \sqrt{h}\right) \right)~.
\end{align}
Hence, the only terms in $-2\partial _{i}\left(\sqrt{-g}A^{i}\right)$ that contain time derivatives of $N$ and $N^{\alpha}$ is the term $-2\partial _{t}\left(N\sqrt{h}A^{t}\right)$, which has the expression
\begin{align}
-2\partial _{t}\left[N\sqrt{h}\left(Kn^{t}+a^{t}\right)\right]=
-2\partial _{t}\left[\sqrt{h}\frac{\partial _{\alpha}N^{\alpha}}{N}
-\frac{1}{N}\partial _{t}\left(\sqrt{h}\right)
+\frac{N^{\alpha}}{N}\partial _{\alpha}\sqrt{h} \right]~.
\end{align}
Thus isolating the time derivatives of $N$ and $N^{\alpha}$ we obtain \ref{Kt}. Let us next calculate the boundary term in \EH action given in \ref{Eq10}, which has the expression $\partial _{i}\left(\sqrt{-g}V^{i}\right)$. Individual components of the boundary term turn out to be
\begin{align}
\sqrt{-g}V^{t}&=\frac{1}{N\sqrt{h}}\partial _{b}\left(-N^{2}hg^{bt}\right)
\nonumber
\\
&=\frac{1}{N\sqrt{h}}\partial _{t}\left(-N^{2}hg^{tt}\right)
+\frac{1}{N\sqrt{h}}\partial _{\alpha}\left(-N^{2}hg^{\alpha t}\right)
\nonumber
\\
&=\frac{1}{N\sqrt{h}}\partial _{t}\left(h\right)
+\frac{1}{N\sqrt{h}}\partial _{\alpha}\left(-hN^{\alpha}\right),
\end{align}
while the spatial component turns out to be
\begin{align}
\sqrt{-g}V^{\alpha}&=\frac{1}{N\sqrt{h}}\partial _{t}\left(-N^{2}hg^{t\alpha}\right)
+\frac{1}{N\sqrt{h}}\partial _{\beta}\left(-N^{2}hg^{\alpha \beta}\right)
\nonumber
\\
&=-\frac{2N^{\alpha}\partial _{t}\sqrt{h}}{N}-\frac{\sqrt{h}\partial _{t}N^{\alpha}}{N}
-\frac{1}{N\sqrt{h}}\partial _{\beta}\left[N^{2}h \left(h^{\beta \alpha}-\frac{N^{\alpha}N^{\beta}}{N^{2}}\right) \right]~.
\end{align}
Hence one arrives at the time derivatives of $N$ and $N^{\alpha}$ in the boundary term as in \ref{Vtd}. Further, from \ref{Vtd} and \ref{Kt}, we finally arrive at \ref{surf_term_diff} used in the main text. Subtraction of these two counter terms confirms that the ADM Lagrangian does not contain any time derivatives of $N$ and $N^{\alpha}$ as emphasized in the text.
\section{Appendix: Derivation of boundary term for null surfaces}\label{PApp_02}

In this appendix we would derive the final expression for boundary term on a null surface as presented in \ref{Final_Eq}. For that we will start with \ref{Eq1} and evaluate each term individually. Let us start with the first divergence term and write it explicitly as:
\begin{align}
Q_{1}=\sqrt{-g}\nabla _{c}\left(\delta u^{c}\right)&=\sqrt{-g}\nabla _{c}\left(P^{c}_{~d}\delta u^{d}\right)
-\sqrt{-g}\nabla _{c}\left(k^{c}\ell _{d}\delta u^{d}\right)
\nonumber
\\
&=\partial _{c} \left(\sqrt{-g}P^{c}_{~d}\delta u^{d}\right)-\sqrt{-g}\nabla _{c}\left(k^{c}\delta \ell ^{2}\right)
\nonumber
\\
&=\partial _{\alpha} \left(\sqrt{-g}P^{\alpha}_{~d}\delta u^{d}\right)-\partial _{c}\left(\sqrt{-g}k^{c}\delta \ell ^{2}\right)~.
\end{align}
The last term can be manipulated as:
\begin{align}
-\partial _{c}\left(\sqrt{-g}k^{c}\delta \ell ^{2}\right)&=-\sqrt{-g}k^{c}\partial _{c}\delta \ell ^{2}
-\delta \ell ^{2}\partial _{c}\left(\sqrt{-g}k^{c}\right)
\nonumber
\\
&=-\delta \left(\sqrt{-g}k^{c}\partial _{c}\ell ^{2} \right)
+\left(k^{c}\partial _{c}\ell ^{2}\right)\delta \sqrt{-g}
\nonumber
\\
&+\sqrt{-g}\delta k^{c}\partial _{c}\ell ^{2}
-\partial _{c}\left( \sqrt{-g}k^{c}\right)\delta \ell ^{2}~.
\end{align}
Hence the first term in \ref{Eq1} takes the following form:
\begin{align}
Q_{1}&=\partial _{\alpha} \left(\sqrt{-g}P^{\alpha}_{~d}\delta u^{d}\right)-\delta \left(\sqrt{-g}k^{c}\partial _{c}\ell ^{2} \right)+\left(k^{c}\partial _{c}\ell ^{2}\right)\delta \sqrt{-g}
\nonumber
\\
&+\sqrt{-g}\delta k^{c}\partial _{c}\ell ^{2}
-\partial _{c}\left( \sqrt{-g}k^{c}\right)\delta \ell ^{2}~.
\end{align}
From which it immediately follows that,
\begin{equation}\label{App_new_01}
\nabla _{c}\delta u^{c}=\frac{1}{\sqrt{-g}}\partial _{\alpha} \left(\sqrt{-g}P^{\alpha}_{~d}\delta u^{d}\right)-\delta \left(k^{c}\partial _{c}\ell ^{2} \right)+\delta k^{c}\partial _{c}\ell ^{2}
-\frac{1}{\sqrt{-g}}\partial _{c}\left( \sqrt{-g}k^{c}\right)\delta \ell ^{2}~,
\end{equation}
while the second term can be expressed as:
\begin{align}
Q_{2}&=-2\delta \left(\sqrt{-g}\nabla _{a}\ell ^{a}\right)
\nonumber
\\
&=-2\delta \left(\sqrt{-g}P^{c}_{~d}\nabla _{c}\ell ^{d}\right)
+\delta \left(\sqrt{-g}k^{c}\partial _{c}\ell ^{2}\right)~.
\end{align}
Using the above trick we can also obtain,
\begin{equation}\label{App_new_02}
-2\delta \left(\nabla _{a}\ell ^{a}\right)=-2\delta \left(P^{c}_{~d}\nabla _{c}\ell ^{d}\right)
+\delta \left(k^{c}\partial _{c}\ell ^{2}\right)~.
\end{equation}
Then writing $\delta \sqrt{-g}=-(1/2)\sqrt{-g}g_{ab}\delta g^{ab}$, the total boundary term takes the form given in \ref{Eq1a}. Let us now concentrate on the null case, where we have $\ell ^{2}:=0$, $k^{2}:=0$ and $\ell .k:=-1$. Then we have $g_{ab}=q_{ab}-\ell _{a}k_{b}-\ell _{b}k_{a}$. We will start with the following expression:
\begin{align}
g_{ab}\delta g^{ab}&=g_{ab}\delta q^{ab}-2\ell _{a}\delta k^{a}-2k_{a}\delta \ell ^{a}
\nonumber
\\
&=q_{ab}\delta q^{ab}-2\ell _{a}k_{b}\delta q^{ab}-2\ell _{b}\delta k^{b}-2k_{a}\delta \ell ^{a}~.
\end{align}
Now we have the following identity:
\begin{align}
\ell _{a}k_{b}\delta q^{ab}&=k_{b}\delta \left(\ell _{a}q^{ab}\right)-k_{b}q^{ab}\delta \ell _{a}
\nonumber
\\
&:=k_{b}\delta \left(\ell ^{b}+\ell ^{2}k^{b}+\lbrace \ell ^{c}k_{c} \rbrace \ell ^{b}\right)
\nonumber
\\
&:=k_{b}\delta \ell ^{b}-\delta (\ell ^{c}k_{c})-k_{b}\delta \ell ^{b}
:=-\delta \left(\ell _{c}k^{c}\right)~.
\end{align}
Thus we arrive at the following expression: 
\begin{align}
g_{ab}\delta g^{ab}:=q_{ab}\delta q^{ab}-2\ell _{b}\delta k^{b}-2k_{a}\delta \ell ^{a}+2\delta \left(\ell _{c}k^{c}\right)~.
\end{align}
Again we have another decomposition:
\begin{align}\label{Eq2}
\nabla _{a}\ell _{b}\delta g^{ab}&:=\left(\Theta _{ab}-\ell _{a}k^{m}\nabla _{m}\ell _{b}-\ell _{b}k^{n}\nabla _{a}\ell _{n}-\ell _{a}\ell _{b}k^{m}k^{n}\nabla _{m}\ell _{n}\right)\delta g^{ab}~.
\end{align}
The first term in the above expression can be simplified, leading to,
\begin{align}
\Theta _{ab}\delta g^{ab}&=\Theta _{ab}\delta q^{ab}-2\Theta _{ab}\ell ^{a}\delta k^{b}-2\Theta _{ab}k^{b}\delta \ell ^{a}
\nonumber
\\
&:=\Theta _{ab}\delta q^{ab}~.
\end{align}
We also have the following two identities given by:
\begin{align}
\ell _{a}\delta g^{ab}&=\delta \ell ^{b}-g^{ab}\delta \ell _{a}~,
\\
\ell _{a}\ell _{b}\delta g^{ab}&=\ell _{a}\delta \ell ^{a}-\ell ^{a}\delta \ell _{a}~.
\end{align}
Hence in total we have:
\begin{align}\label{App_new_03}
\nabla _{a}\ell _{b}\delta g^{ab}&:=\Theta _{ab}\delta q^{ab}
-\left(k^{m}\nabla _{m}\ell _{b}\right)\ell _{a}\delta g^{ab}
-\left(k^{n}\nabla _{a}\ell _{n}\right)\ell _{b}\delta g^{ab}
\nonumber
\\
&-\left(k^{m}k^{n}\nabla _{m}\ell _{n}\right)\ell _{a}\ell _{b}\delta g^{ab}
\nonumber
\\
&:=\Theta _{ab}\delta q^{ab}-\left(k^{m}\nabla _{m}\ell _{b}\right)\delta \ell ^{b}
+\left(k^{m}\nabla _{m}\ell ^{a}\right)\delta \ell _{a}
-\left(k^{n}\nabla _{a}\ell _{n}\right)\delta \ell ^{a}
\nonumber
\\
&+\left(k^{n}\nabla ^{b}\ell _{n}\right) \delta \ell _{b}
-\left(k^{m}k^{n}\nabla _{m}\ell _{n}\right)\left(\ell _{a}\delta \ell ^{a}-\ell ^{a}\delta \ell _{a}\right)
\nonumber
\\
&:=\Theta _{ab}\delta q^{ab}-\left\lbrace k^{m}\nabla _{m}\ell _{a}+k^{n}\nabla _{a}\ell _{n}
+\left(k^{m}k^{n}\nabla _{m}\ell _{n}\right)\ell _{a} \right\rbrace \delta \ell ^{a}
\nonumber
\\
&+\left\lbrace k^{m}\nabla _{m}\ell ^{a}+k^{n}\nabla ^{a}\ell _{n}
+\left(k^{m}k^{n}\nabla _{m}\ell _{n}\right)\ell ^{a} \right\rbrace \delta \ell _{a}~.
\end{align}
Thus the last term in \ref{Eq1a} takes the following form:
\begin{align}
\Bigg[\nabla _{a}\ell _{b}&-g_{ab}\left(P^{c}_{~d}\nabla _{c}\ell ^{d}\right)\Bigg]\delta g^{ab}:=
\Theta _{ab}\delta q^{ab}
\nonumber
\\
&-\left\lbrace k^{m}\nabla _{m}\ell _{a}+k^{n}\nabla _{a}\ell _{n}
+\left(k^{m}k^{n}\nabla _{m}\ell _{n}\right)\ell _{a} \right\rbrace \delta \ell ^{a}
\nonumber
\\
&+\left\lbrace k^{m}\nabla _{m}\ell ^{a}+k^{n}\nabla ^{a}\ell _{n}
+\left(k^{m}k^{n}\nabla _{m}\ell _{n}\right)\ell ^{a} \right\rbrace \delta \ell _{a}
\nonumber
\\
&-\left(P^{c}_{~d}\nabla _{c}\ell ^{d}\right)\left[q_{ab}\delta q^{ab}-2\ell _{b}\delta k^{b}-2k_{a}\delta \ell ^{a}
+2\delta(\ell _{c}k^{c})\right]
\nonumber
\\
&:=\left[\Theta _{ab}-\left(P^{c}_{~d}\nabla _{c}\ell ^{d}\right)q_{ab}\right]\delta q^{ab}
\nonumber
\\
&-\left\lbrace k^{m}\nabla _{m}\ell _{a}+k^{n}\nabla _{a}\ell _{n}
+\left(k^{m}k^{n}\nabla _{m}\ell _{n}\right)\ell _{a}-2\left(P^{c}_{~d}\nabla _{c}\ell ^{d}\right)k_{a} \right\rbrace \delta \ell ^{a}
\nonumber
\\
&+\left\lbrace k^{m}\nabla _{m}\ell ^{a}+k^{n}\nabla ^{a}\ell _{n}
+\left(k^{m}k^{n}\nabla _{m}\ell _{n}\right)\ell ^{a}-2\left(P^{c}_{~d}\nabla _{c}\ell ^{d}\right)k^{a}  \right\rbrace \delta \ell _{a}~.
\end{align}
Thus combining all the expressions we finally arrive at \ref{Final_Eq}. Along identical lines, using \ref{App_new_01}, \ref{App_new_02} and \ref{App_new_03} one arrives at \ref{Final_Eq_Sim}. 

If one assumes $\delta \ell _{a}=0$, then from the condition $\delta (\ell_{a}k^{a})=0$, it follows that $\delta k^{a}$ has to lie on the hypersurface and hence $\delta k^{c}\partial _{c}\ell ^{2}=0$. Also, from the condition $\nabla _{a}\ell _{b}=\nabla _{b}\ell _{a}$, we obtain, $k^{m}\nabla _{m}\ell _{a}=k^{n}\nabla _{a}\ell _{n}$. These results are used to arrive at \ref{Derived_01}.

While if, $\delta \ell _{a}=(\delta \ln A)\ell _{a}$, one obtains from $\delta (\ell _{a}k^{a})=0$ the following result: $\ell _{a}\delta k^{a}=\delta \ln A$. Thus $\delta k^{a}=-(\delta \ln A)k^{a}+\textrm{surface components}$. Hence, $\delta k^{c}\partial _{c}\ell ^{2}=-(\delta \ln A)k^{c}\partial _{c}\ell ^{2}$. Along identical lines, from $\delta (\ell^{2})=0$, we arrive at, $\ell _{a}\delta \ell ^{a}=0$. Further,
\begin{align}
\Big\lbrace &k^{m}\nabla _{m}\ell ^{a}+k^{n}\nabla ^{a}\ell _{n}
+\left(k^{m}k^{n}\nabla _{m}\ell _{n}\right)\ell ^{a} -2\left(P^{m}_{~n}\nabla _{m}\ell ^{n}\right)k^{a}+\nabla ^{a}\ln A \Big\rbrace \delta \ell _{a}
\nonumber
\\
&\left\lbrace \partial _{c}\ell ^{2}\right\rbrace \delta k^{c}:=\left(\delta \ln A \right)\ell _{a}\Big\lbrace k^{m}\nabla _{m}\ell ^{a}+k^{n}\nabla ^{a}\ell _{n}
\nonumber
\\
&+\left(k^{m}k^{n}\nabla _{m}\ell _{n}\right)\ell ^{a} -2\left(\Theta +\kappa \right)k^{a}+\nabla ^{a}\ln A \Big\rbrace 
+\left\lbrace \partial _{c}\ell ^{2}\right\rbrace \delta k^{c}
\nonumber
\\
&:=2\left(\Theta +\kappa \right)\left(\delta \ln A \right)+\left(\delta \ln A \right)\left(\ell ^{a}\nabla _{a}\ln A\right)
-\kappa \left(\delta \ln A \right)-\frac{1}{2}k^{c}\partial _{c}\ell ^{2}\left(\delta \ln A \right)
\nonumber
\\
&:=2\left(\Theta +\kappa \right)\left(\delta \ln A \right)~,
\end{align}
where in the last but one line we have used the result that, $\kappa =\ell ^{a}\nabla _{a}\ln A-(1/2)k^{c}\nabla _{c}\ell ^{2}$. Substitution of the above relation as well as the results obtained in the previous paragraphs in \ref{Final_Eq_A} leads to \ref{Derived_02}.

\bibliography{Gravity_1_full,Gravity_2_partial,My_References}

\bibliographystyle{./utphys1}
\end{document}